# Strain Mapping of Two-Dimensional Heterostructures with Sub-Picometer Precision

Yimo Han[1]*, Kayla Nguyen[3], Michael Cao[1], Paul Cueva[1], Saien Xie[1,2], Mark W. Tate[4], Prafull Purohit[4], Sol M. Gruner[4,5,6,7], Jiwoong Park[3], David A. Muller[1,6]*

[1.] School of Applied and Engineering Physics, Cornell University, Ithaca, NY 14853, USA

[2.] Department of Chemistry, Institute for Molecular Engineering, and James Franck Institute, University of Chicago, Chicago, IL 60637, USA

[3.] Chemistry and Chemical Biology Department, Cornell University, Ithaca, NY 14853, USA

[4.] Laboratory of Atomic and Solid State Physics, Cornell University, Ithaca, NY 14853, USA

[5.] Physics Department, Cornell University, Ithaca, NY 14853, USA

[6.] Kavli Institute at Cornell for Nanoscale Science, Ithaca, NY 14853, USA

[7.] Cornell High Energy Synchrotron Source, Cornell University, Ithaca, NY 14853, USA

*Correspondence to: david.a.muller@cornell.edu, yh542@cornell.edu

**Next-generation, atomically thin devices require in-plane, one-dimensional heterojunctions to electrically connect different two-dimensional (2D) materials. However, the lattice mismatch between most 2D materials leads to unavoidable strain, dislocations, or ripples, which can strongly affect their mechanical, optical, and electronic properties. We have developed an approach to map 2D heterojunction lattice and strain profiles with sub-picometer precision and to identify dislocations and out-of-plane ripples. We collected diffraction patterns from a focused electron beam for each real-space scan position with a high-speed, high dynamic range, momentum-resolved detector – the electron microscope pixel array detector (EMPAD). The resulting four-dimensional (4D) phase space datasets contain the full spatially resolved lattice information of the sample. By using this**

**technique on tungsten disulfide (WS$_2$) and tungsten diselenide (WSe$_2$) lateral heterostructures, we have mapped lattice distortions with 0.3 pm precision across multi-micron fields of view and simultaneously observed the dislocations and ripples responsible for strain relaxation in 2D laterally-epitaxial structures.**

Strain fields and dislocations play an important role in determining the mechanical and electronic properties of crystalline materials. Conventional transmission electron microscopy (TEM) can identify strain fields and dislocations in bulk materials using the diffraction contrast from two-beam analysis[1] or weak-beam dark field imaging[2]. However, given the rod-like nature of diffraction peaks normal to the film, these methods are very difficult to implement in 2D materials, which are confined to atomic dimensions in the direction of the beam propagation. While geometric phase analysis (GPA)[3-6] on atomic-resolution images can provide strain maps for 2D materials, the field-of-view is limited to a few tens of nanometers by the need to resolve every atom. When grains or heterostructures of 2D materials[7-10] laterally reach the more typical micron scales, measuring strain and dislocations at atomic resolution using GPA becomes impractical. Conversely, nanobeam diffraction (NBD) combined with scanning TEM (STEM)[11-15] effectively decouples the spatial resolution from the strain mapping precision, allowing for high-precision strain measurements across a larger sample area. However, this approach on 2D materials has historically been limited by the speed and dynamic range of the existing detectors, as 2D materials are sensitive to the electron beam and weak scatterers.

To overcome these issues, we developed a method to map the strain and identify dislocations in 2D crystals using an electron microscope pixel array detector (EMPAD)[16] designed at Cornell. Due to its high-speed, high dynamic range, and high sensitivity, the scanning NBD can be achieved within minutes with no noticeable damage to 2D samples, ultimately providing sub-picometer precision strain mapping over length scales, ranging from angstroms to many micrometers. The EMPAD operates at a range of accelerating voltages from 20 kV to 300 kV (The work presented here was conducted at 80 kV). Its single-electron sensitivity allows for quantitative analysis of diffraction from a single atom[17,18], which is highly advantageous for studying 2D materials only one- to three-atoms thick. Moreover, the EMPAD's high dynamic range enables collection of all transmitted electrons at small convergence angles with the primary beam unsaturated and diffracted beams clearly resolved (Fig. 1b), as demonstrated by integrating the center beam (and one diffracted spot) to plot the virtual bright field (and dark field) images, as shown in Fig. 1c and 1d.

The samples we examined were laterally stitched epitaxial monolayer transition metal dichalcogenide (TMD) heterojunctions, which are synthesized through metal-organic chemical vapor deposition (MOCVD)[19,20]. From the EMPAD's 4D dataset of a sample on 20 nm $SiO_2$ TEM grids, we extracted the ADF-STEM signal (Fig. 2a) by integrating the diffraction patterns masked by a virtual ADF detector with an inner angle of 50 mrad. The ADF-STEM image provides little contrast difference between $WS_2$ and $WSe_2$, as the heavy tungsten atoms dominate the contrast. In comparison, the lattice constant map (Fig. 2b) calculated by measuring shifts in the reciprocal lattice vectors clearly distinguishes the two materials, which contain nanometer-sharp interfaces (inset of Fig. 2b) (see SI#1 and #2 for calculation details). From a histogram of lattice constant measurment (Fig. 2c), we extracted the mean values as the statistically averaged lattice constant for $WS_2$ (3.182 ± 0.0005 Å) and $WSe_2$ (3.282 ± 0.001 Å), indicating a 3.1% lattice mismatch (fully relaxed films have 4.5% lattice mismatch[21]). The histogram (inset of Fig. 2c) from a flat region (gray box in Fig. 2b) indicates this method has a precision higher than 0.3 pm, with local sample distortions placing an upper limit on the spread. The sub-picometer precision of the lattice constant mapping relies on high angular resolution when we measure the centers of the diffraction spots. For 2D materials, the center of mass (CoM) is an efficient approach to achieve a high angular resolution in the diffraction space for strain mapping. The calculation details, optimization of pixel sampling, and error analysis are provided in supplementary information #1.

From the CoMs of all diffraction spots, we can also extract the diffraction vectors $g_{i\ (i=1,2)}$ (Fig. S4), i.e. the reciprocal lattice vectors, which can be used to map the strain and rotation (see details in SI #3). The x-direction uniaxial strain ($\varepsilon_{xx}$) map (Fig. 3a) shows clear differences between the WS$_2$ and WSe$_2$ as well as small local variations, indicating the film is largely relaxed, but not completely. This is expected since the width of WS$_2$ is far beyond the critical thickness (a few nanometers for TMD materials), as narrow heterojunctions below the critical thickness remain coherent and exhibit uniaxial strain parallel to the interface (Fig. S5). To see the strain details, we plot the histograms of the x-direction strain map, where the WS$_2$ peak in Fig. 3b fits two Gaussian peaks, corresponding to the unstrained (outer edges) and strained (interfaces) parts of the WS$_2$ lattice in Fig. 3a. From a relatively flat WS$_2$ region (gray box in Fig. 3a), we determined that the precision of our technique is at least 0.18%, as given by the spread of the histogram in the insert of Fig. 3b.

The rotation map (Fig. 3c) displays periodic misfit dislocations as dipole fields located at the interface between WS$_2$ and WSe$_2$. The misfit dislocations along the interface contribute to release the lattice strain. However, the observed misfit dislocation spacing (~100 nm) is much larger than the spacing required to fully relax the lattice strain caused by the 3.1% lattice mismatch between WS$_2$ and WSe$_2$ (~10 nm). We note an internal periodic strain field in the outer WS$_2$ region in the rotation map (Fig. 3c). Analogous to bulk epitaxy, which forms periodic ripples within the top layers of the thin films[22], the periodic strain fields lower the elastic strain energy in 2D heterojunctions and create intrinsic dislocations inside WS$_2$, as depicted in the magnified rotation map (Fig. 3d). However, the long wavelength of the strain field (in microns) implies that it plays a minor role in releasing strain. A more significant contribution for releasing strain is that 2D materials can also buckle up and form out-of-plane ripples without introducing in-plane lattice distortions. We observed that the WSe$_2$ forms this type of ripples, which is the origin of WSe$_2$'s broader peak in Fig. 3b. These out-of-plane ripples contribute to relax the lattice strain dramatically[20,23].

To accurately identify the out-of-plane ripples and quantitatively map their orientations, we developed a novel approach, using the EMPAD 4D datasets. Due to the cone-shaped diffracting beam from the microscopic corrugation of 2D materials[24], the tilt of the 2D materials results in elliptical diffraction spots, as illustrated schematically in Fig. 4a.

Here, we mapped the tilted regions (i.e. the ripples) by measuring the relative broadening of the diffraction spots compared to flat regions. We defined the complex ripple measure as $R = A + Be^{i2\pi/3} + Ce^{i4\pi/3}$, where A, B, and C are the characteristic sizes (as measured by second moments) of the corresponding spots in Fig. 4a. As illustrated in the phase plot of the R maps, the ripples form along different orientations, and only appear inside $WSe_2$ (Fig. 4b). In addition, the ripples prefer to form perpendicular to the interface between $WS_2$ and $WSe_2$ on the $WSe_2$ side, thus releasing the lattice strain, as shown in Fig. 4c. The calculation details can be found in supplementary information #4.

So far, we mapped the lattice constant, strain, rotation, and ripples in the lateral heterostructure and identified uniaxial strain, dislocations, and ripples. All the lattice information is recorded within a single EMPAD 4D dataset. In electron microscopy, principle component analysis (PCA) has been widely used for multi-dimensional datasets, such as electron energy loss spectra (EELS)[25,26]. In EMPAD 4D datasets, PCA as a screening tool provides a statistical approach to immediately obtain a hierarchy of information by extracting the linearly uncorrelated patterns that contain the majority of the variance in the data (SI #5). For our 4D data of the $WS_2$-$WSe_2$ multi-junctions, we blocked the center beam for PCA to avoid any dominating features inside the center beam. The principle components of the diffraction patterns (Fig. 5c-h) and their corresponding weighting maps (Fig. 5i-l) capture the highest variance features in the 4D dataset (more maps in Fig. S7). By comparing the maps generated using the former method (Fig. 5a-d), we can relate the second to fifth principle components to the filtered dark field (DF) of the multi-junction (Fig. 5e, i), the rotation of the lattice (Fig. 5f, j), the lattice constant difference (Fig. 5g, k), and the local ripples in $WSe_2$ region (Fig. 5h, l), respectively. PCA offers a swift and facile method to analyze the 4D EMPAD data and quickly highlights the largest changes for our future studies.

In conclusion, we developed an approach to map lattice constant, strain, dislocations, and out-of-plane ripples with high precision on all relevant length scales. All lattice information can be extracted from the 4D data which requires only a single fast scan, effectively reducing the electron dose in 2D materials. Moreover, the accuracy of CoM measurements allows us to map the lattice constant and strain with precisions greater than 0.3 pm and 0.18% respectively at $WS_2$-$WSe_2$ lateral heterojunctions. In addition, we observed that the lattice strain is released mainly by misfit dislocations and out-of-plane ripples, both of which play an important role in the mechanical and electrical properties of 2D materials. This method will be crucial for studying the lattice distortions in 2D materials and their atomically thin circuitry by identifying the lattice strain and dislocations and offering essential feedback to material syntheses.

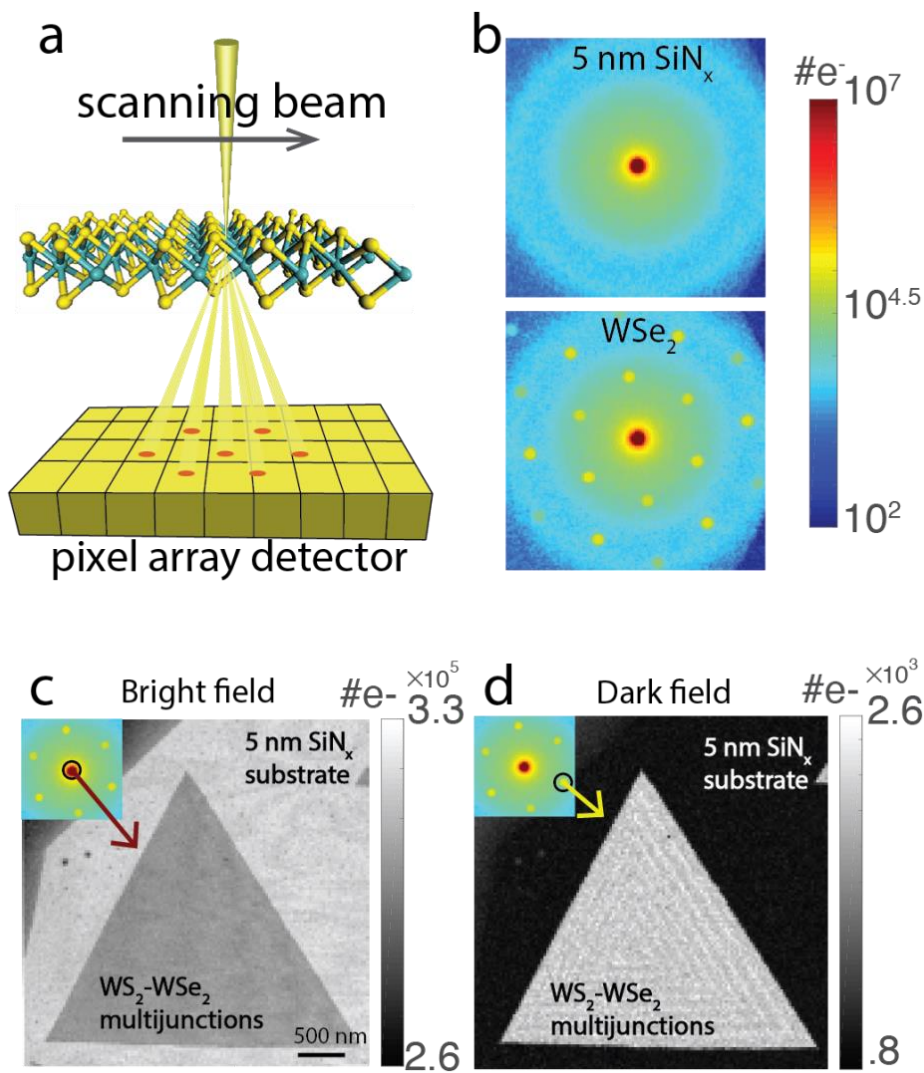

**Figure 1 | EMPAD imaging. a,** Schematic of the EMPAD operation, where a full diffraction pattern, including the unsaturated primary beam, is recorded at each scan position. **b,** Diffraction images taken by EMPAD. The top panel shows the diffraction image of a 5 nm $SiN_x$ film, while the bottom panel displays the diffraction pattern of a $WSe_2$ monolayer located on the 5 nm $SiN_x$ film. **c** and **d** show the virtual bright field and filtered dark field images obtained by integrating the central and the labeled diffracted beam, respectively, as indicated on their top left sections.

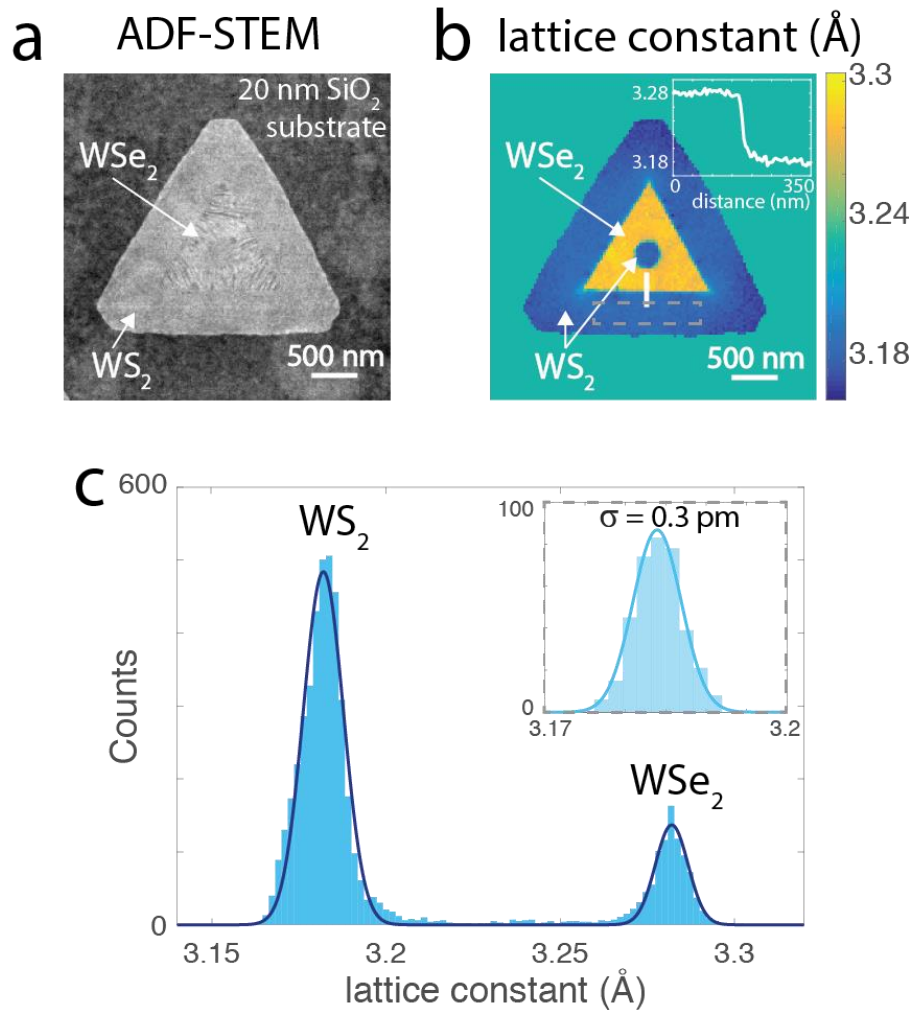

**Figure 2 | Lattice constant map. a,** Annular dark field (ADF) image extracted from the EMPAD 4D data on a wide $WS_2$-$WSe_2$ lateral heterojunction. The inner detector angle is 50 mrad. **b,** Lattice constant map of micron-sized triangles. The inset displays a line profile across the interface between $WSe_2$ and $WS_2$. **c,** Lattice constant histogram from **b**. The inset is the histogram from a flat region (gray box in **b**), indicating a precision of at least ~0.3 pm.

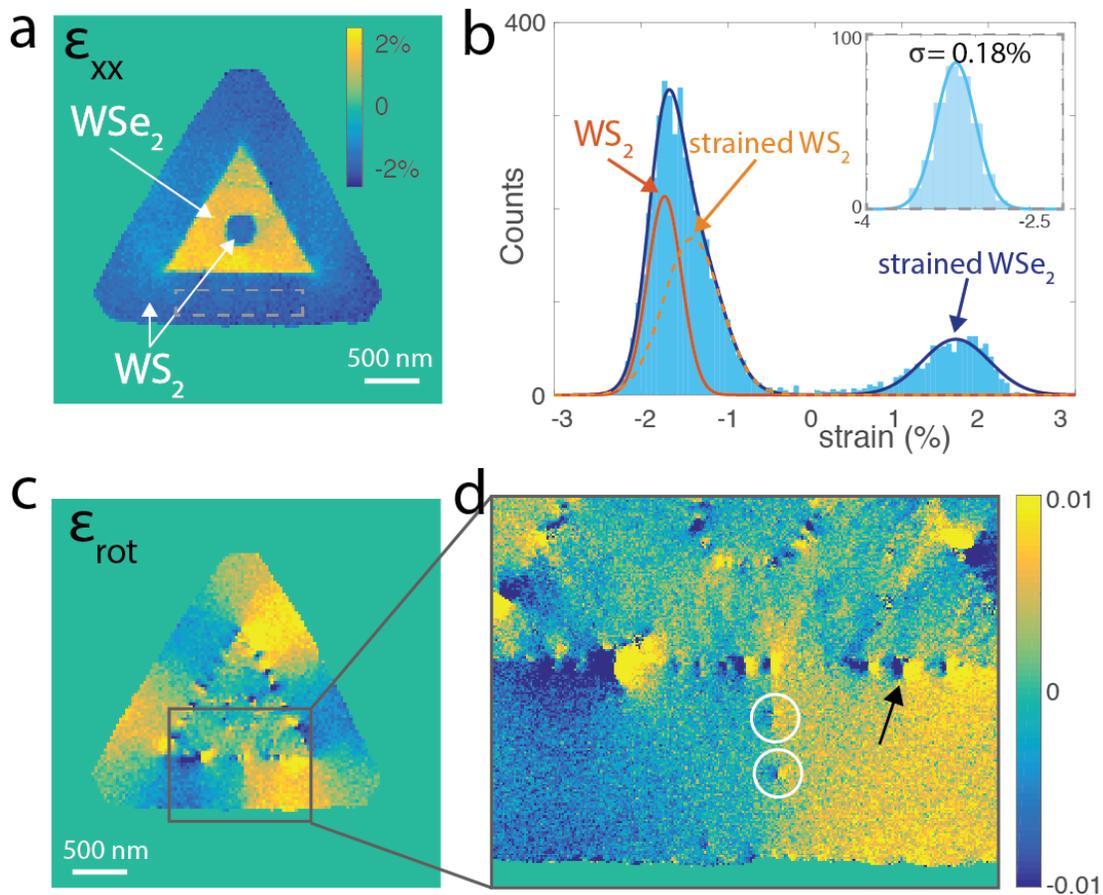

**Figure 3 | Strain maps. a,** Uniaxial strain map showing most of the strain has been released. **b,** Strain histogram from **a**. The $WS_2$ peak fits two Gaussians that correspond to unstrained and strained regions. The inset shows the histogram from a flat region (gray box in **a**) indicating a precision better than ~ 0.18%. **c,d,** The rotation map displaying periodic misfit dislocations that contribute to relaxing the lattice strain at the $WS_2$-$WSe_2$ junction. The internal strain in $WS_2$ results in a few dislocations inside $WS_2$, which are indicated by the white circles.

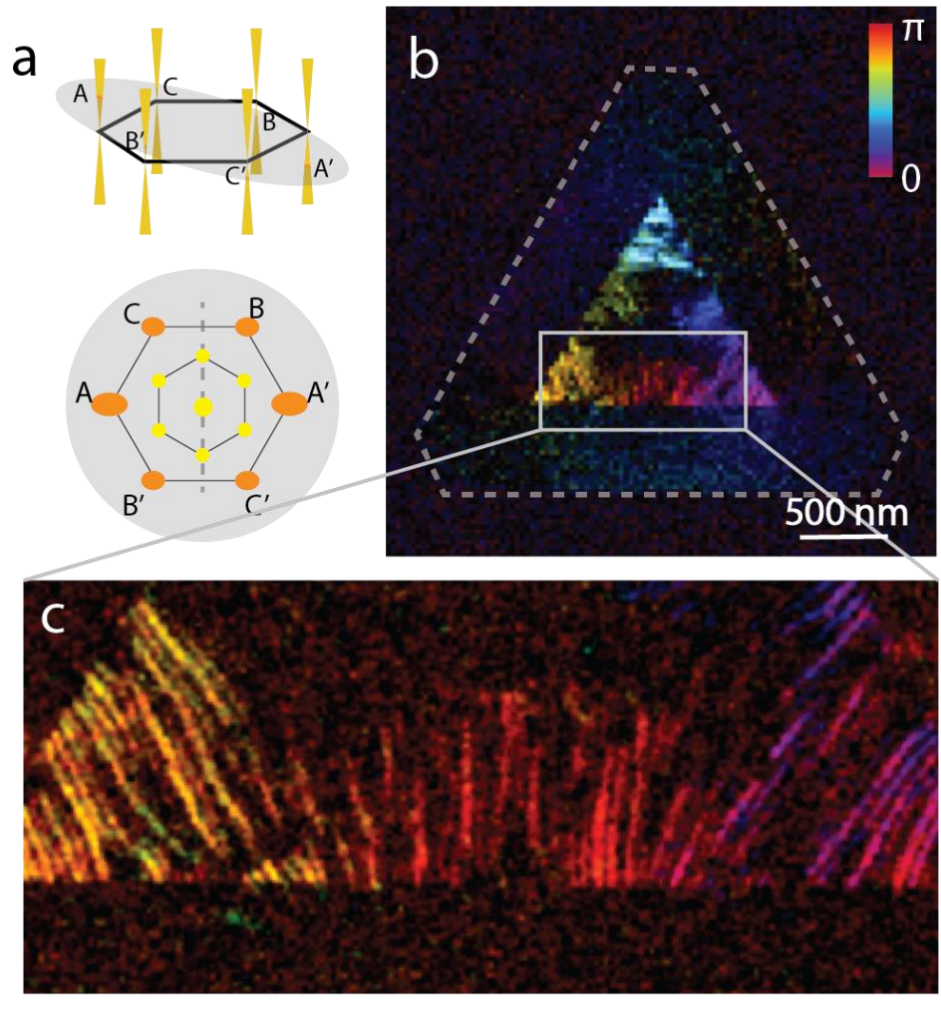

**Figure 4 | Map of out-of-plane ripples. a,** Schematics showing that the tilt causes the ellipsoidal broadening of the diffraction spots due to the corrugation of the 2D materials. The broadening of the spots was measured through the second moment measurement. **b,** The phase plot of relative broadening shows the oriented ripple map where the WSe$_2$ film forms out-of-plane ripples to release the strain, while WS$_2$ is flat. **c,** Magnified ripple map showing a nanoscale ripple array along the junction.

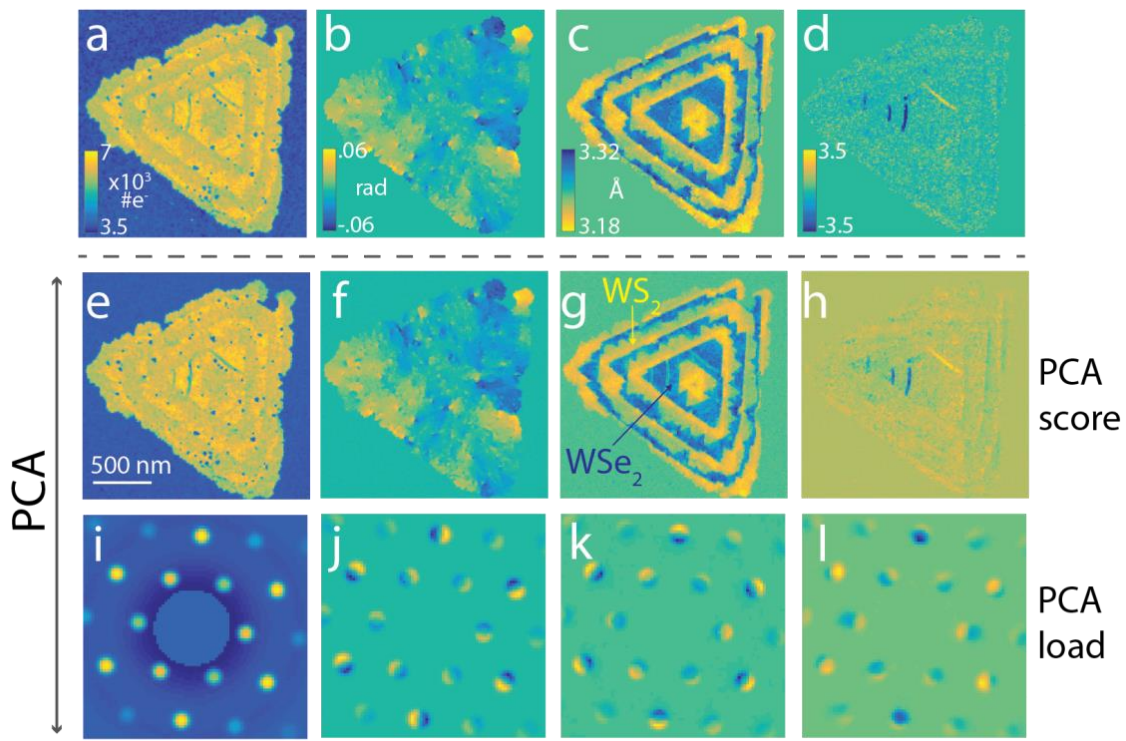

**Figure 5 | Extracted data vs PCA on EMPAD 4D data. a** Filtered DF, **b** rotation map, **c** lattice constant map, and **d** ripple map. **e-h,** The PCA scores of the second to fifth principle components of the EMPAD 4D dataset and **i-l** their corresponding PCA loads. **e,i** show the crystalline structure, **f,j** depict the lattice rotation, **g,k** represent the lattice constant difference, and **h,l** display the local ripples.

**Corresponding Author**

*David A. Muller (david.a.muller@cornell.edu), Yimo Han (yh542@cornell.edu)

**Acknowledgements**

This work was primarily supported by the Cornell Center for Materials Research with funding from the NSF MRSEC program (DMR-1719875). MC was supported by DOD-MURI (Grant No. FA9550-16-1-0031). Detector development at Cornell is supported by DOE award DE-SC0017631 to SMG. We thank Mervin Zhao, Megan Holtz, Zhen Chen, Mengnan Zhao, Gabriela Correa, Lei Wang for helpful discussions. We thank Mariena Ramos and Lena Kourkoutis for help with electron microscopes.

**Author contributions**

The project was conceived and designed by Y.H. under the supervision of D.A.M.. The experimental results and data analysis were obtained by Y.H., with help from K.N., M.C., and P.C.. The samples were grown by S.X. under the supervision of J.P.. The EMPAD was developed by M.W.T., P.P., K.N., and M.C. under the supervision of S.M.G. and D.A.M..

# Supporting Information for Strain Mapping of Two-Dimensional Heterostructures with Sub-Picometer Precision

Yimo Han[1]*, Kayla Nguyen[3], Michael Cao[1], Paul Cueva[1], Saien Xie[1,2], Mark W. Tate[4], Prafull Purohit[4], Sol M. Gruner[4,5,6,7], Jiwoong Park[3], David A. Muller[1,6]*

[1.] School of Applied and Engineering Physics, Cornell University, Ithaca, NY 14853, USA

[2.] Department of Chemistry, Institute for Molecular Engineering, and James Franck Institute, University of Chicago, Chicago, IL 60637, USA

[3.] Chemistry and Chemical Biology Department, Cornell University, Ithaca, NY 14853, USA

[4.] Laboratory of Atomic and Solid State Physics, Cornell University, Ithaca, NY 14853, USA

[5.] Physics Department, Cornell University, Ithaca, NY 14853, USA

[6.] Kavli Institute at Cornell for Nanoscale Science, Ithaca, NY 14853, USA

[7.] Cornell High Energy Synchrotron Source, Cornell University, Ithaca, NY 14853, USA

*Correspondence to: david.a.muller@cornell.edu, yh542@cornell.edu

**Contents:**

## 1: Measuring the center of mass (CoM)

CoM calculation

The CoM is calculated from the diffraction pattern, $I(\vec{p})$, using the following equation:

$$\langle \vec{p} \rangle = \int \vec{p}\, I(\vec{p}) d\vec{p} \tag{1}$$

where $\vec{p}$ is the momentum in the diffraction space.

Fig. S1 shows the definition of our masks (green circles) that were applied to the diffraction disks and the center disk when we calculated their CoMs. The green dots label the calculated CoMs of all disks. Fig. S1c shows that the diffraction disk span across ~6 pixels in diameter. The mask with a diameter of 12 pixels is aligned to the disk with 1/3-pixel resolution by eye. We attempted aligning the mask as well as possible and, in fact, we achieved that the center of the mask is close to the measured CoM in Fig. S1d. In addition, measuring the centers for the EMPAD 4D data, generally a few gigabytes in size, requires a fast-computational algorithm. The CoM calculation is an *O(n)* algorithm (see code below), where n can usually be only tens of pixels for each diffraction disk. As a conclusion, CoM is a high-efficiency approach for measuring centers for the EMPAD 4D datasets.

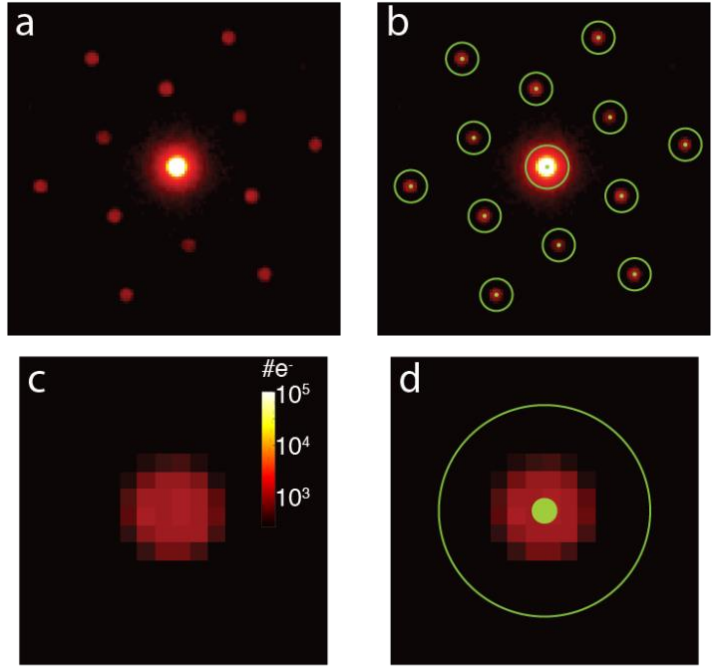

**Figure S1 | CoM measurements. a,** A diffraction pattern of WS$_2$ on a 5 nm SiN$_x$ window taken by EMPAD. **b,** The diffraction pattern overlaid by the masks (green circles) and their CoMs (green dots). We manually placed the masks and aligned them to the diffraction spots by eye. **c,d,** A magnified diffraction disk (c) and its overlay with the mask and the CoM (d).

The accuracy of CoM measurements

      CoM has the advantage of speed and simplicity compared to more elaborate curve fitting procedures. The CoM provides sufficiently accurate centers of diffraction patterns of 2D materials mainly because of the following two reasons: 1) the rod-like nature of diffraction patterns of 2D materials and 2) the high dynamic range of our EMPAD, which counts all transmitted electrons. The errors for CoM measurements come from the unavoidable Poisson noise from the detector.

      Here, we discuss how Poisson noise affects the CoM. For each pixel in the EMPAD, the Poisson noise is proportional to $\sqrt{N}$, where N is the number of electrons hitting that pixel. By calculating the error propagation in Equation (1), we achieved that $\delta CoM = \sqrt{\frac{\langle p^2 \rangle}{N}} \propto \frac{1}{\sqrt{N}}$, where $\delta CoM$ is the absolute error of CoM, and $\langle p^2 \rangle$ is the

second moment defined in Equation (5). This result shows that increasing the electron dose, i.e. using a high beam current, can effectively reduce the error caused by Poisson noise. However, there is a trade-off between high beam current for more accurate CoM and low beam current to avoid electron beam damage in 2D materials. Experimentally, we were working on the regime that the number of electrons in each diffraction spot are on the order of $10^4$ to $10^5$.

Moreover, $\delta CoM$ also depends on the number of pixels the disk spans across (the diameter of the disk in the unit of pixels, as shown in Fig. S2a). We simulated $\delta CoM$ for different disk diameters by averaging the errors from 1000 diffraction patterns with Poisson noise, and plotted the results in Fig. S2b. The results show the absolute error ($\delta CoM$) is proportional to the disk diameter for a given dose.

Designing number of pixels in the detector

The parameter determines the angular resolution is the percentage error, which is the ratio between the absolute error ($\delta CoM$) and the k vector length – $\delta CoM/k$. We note that k, as well as the disk diameter, changes correspondingly as we change the number of pixels in the detector (or change the camera length to magnify the diffraction patterns), as shown in Fig. S2c. In addition, k is proportional to the disk diameter. Here, we ignored a constant scaling prefactor, assumed k equals to the disk diameter, and plotted the percentage errors in Fig. S2d. The percentage errors are close to a constant with small increment at the smaller disk diameters, especially for the low-dose case. For example, if each diffraction disk only contains 10 electrons (blue curve in Fig. S2d), we would choose 10-15 pixels for a disk diameter for reasonable angular resolution. For doses larger than 1000 electrons per disk (we worked at $10^4$ to $10^5$) the optimized disk diameters will be any one larger than 5 pixels. Above that, the errors stay constant, indicating that we do not benefit from designing more pixels in the detector or magnifying the diffraction patterns.

Optimizing convergence angle

For diffraction disk at a fixed k and dose, we will have a smaller percentage error if we have smaller convergence angle (as shown in Fig. S2e and f). Although smaller convergence angles (less pixels) are preferred, the lower bound is a 2x2-pixel-sized diffraction disk, which is similar to a differential phase contrast (DPC) detector. In addition, there is also a tradeoff between the angular resolution and spatial resolution: $d_0 = 1.22\frac{\lambda}{\theta}$, where $d_0$ is the probe size, $\lambda$ is the wavelength of the electrons, and $\theta$ is the convergence angle of the electron beam. In our case, we used a 0.5 mrad convergence angle, where the diffraction disks span several of pixels on the EMPAD, resulting in a 5 nm probe size. Different probe sizes can be selected for the length scale appropriate for the sample, with large probes desirable at large length scales to avoid under sampling.

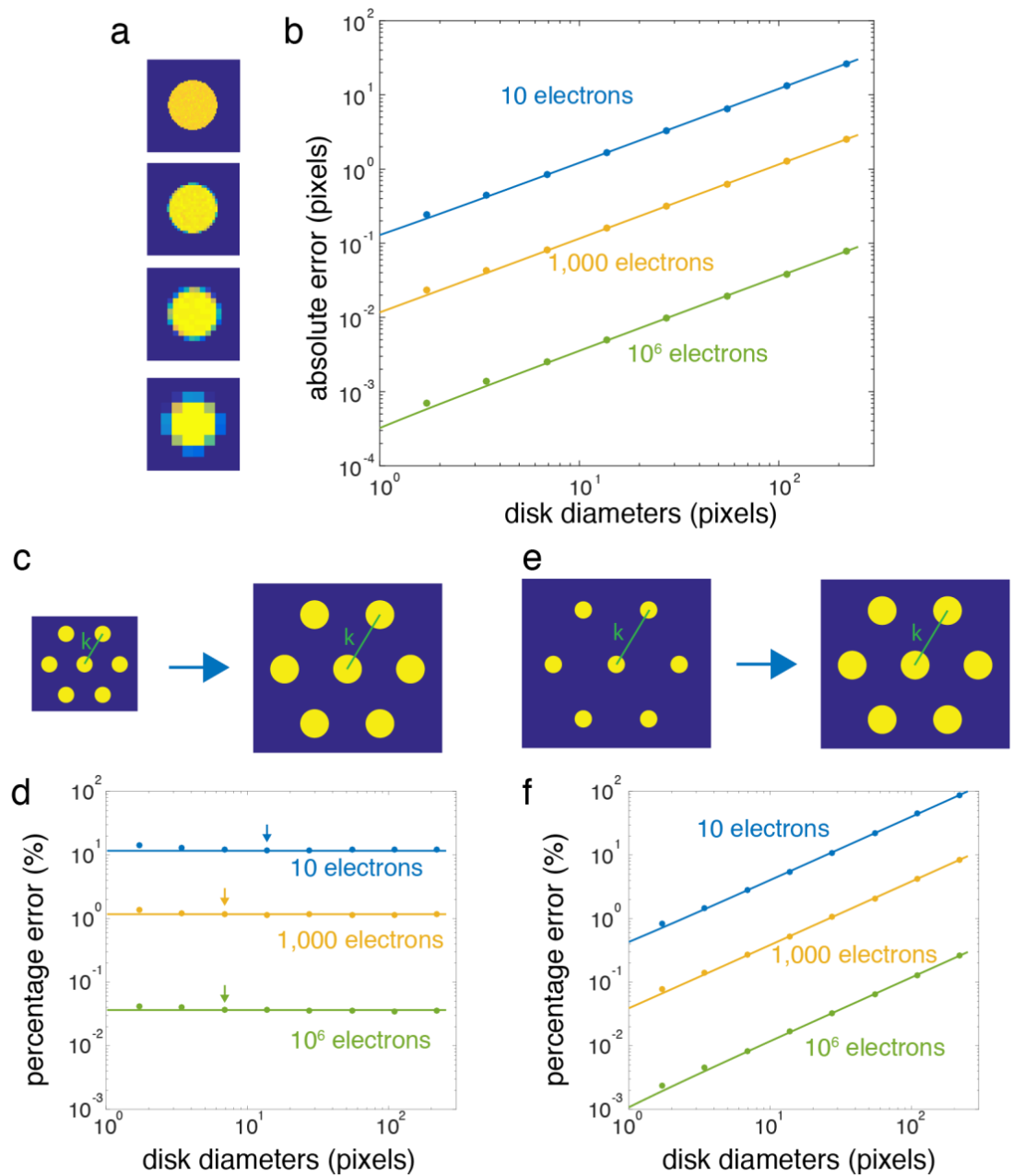

**Figure S2 | Accuracy of CoM measurements. a,** Simulated diffraction patterns of different diameters (in pixels) with Poisson noise. **b,** Absolute errors of CoM ($\delta CoM$) proportional to the disk diameters. **c,** Schematics of diffraction patterns when we increase the number of pixels in the detector (or magnify the diffraction patterns by changing the camera length). **d,** Percentage error plot ($\delta CoM/k$) under the situation described in (c), with arrows indicating the lower bounds of the optimized disk

diameters. (We used k = disk diameter.) **e,** Schematics of diffraction patterns when we change the convergence angle. **f,** Percentage error plot ($\delta CoM/k$) for cases in (e), showing that reducing the convergence angle will reduce the error and improve the angular resolution dramatically. (We used k = 30 pixels.)

## 2: Mapping lattice constant

To calculate the lattice constants from a single diffraction pattern, we averaged the distances between diffracting beams and the center beam, $d_1$ to $d_6$, as shown in Fig. S3a. The averaged lattice constant is:

$$a_{ave} = \frac{6a_0 d_0}{\sum_{i=1}^{6} d_i} \tag{2}$$

where $a_0$ and $d_0$ are the calibrated ones from a referenced region. We used flat $WS_2$ as the reference.

In STEM, EMPAD acquires diffraction patterns at each scan position with 1.86 ms/frame (1 ms exposure time and 0.86 ms readout time) – so a 4D data (x and y in real space and $k_x$ and $k_y$ in momentum space) at 256×256 scan points can be reached in about two minutes. Using the 4D dataset, we can map the lattice constant throughout the entire sample.

Although there are strain and tilt at some regions in the sample, due to the averaging of the six spots in different directions, the strain and tilt effects are negligible. Fig. S3b and c show the schematics of how strain and tilt affect the diffraction pattern. The calculation below describes that the strain and tilt are higher order effects in the lattice constant calculation:

$$a'_{ave} \cong \frac{3a_0 d_0}{d_1(1+\varepsilon') + 2d_2\left(1 - \frac{\sqrt{3}}{2}\nu\varepsilon'\right)} \tag{3}$$

where $\nu$ is the Poisson's ratio (0.25 for $WS_2$) and $\varepsilon'$ is a small uniaxial strain (we used compressive strain here).

$$a''_{ave} \cong \frac{3a_0 d_0}{d_1\left(1 + \frac{\theta^2}{2}\right) + 2d_2\left(1 + \frac{\theta^2}{8}\right)} \tag{4}$$

where $\theta$ is the small tilt angle.

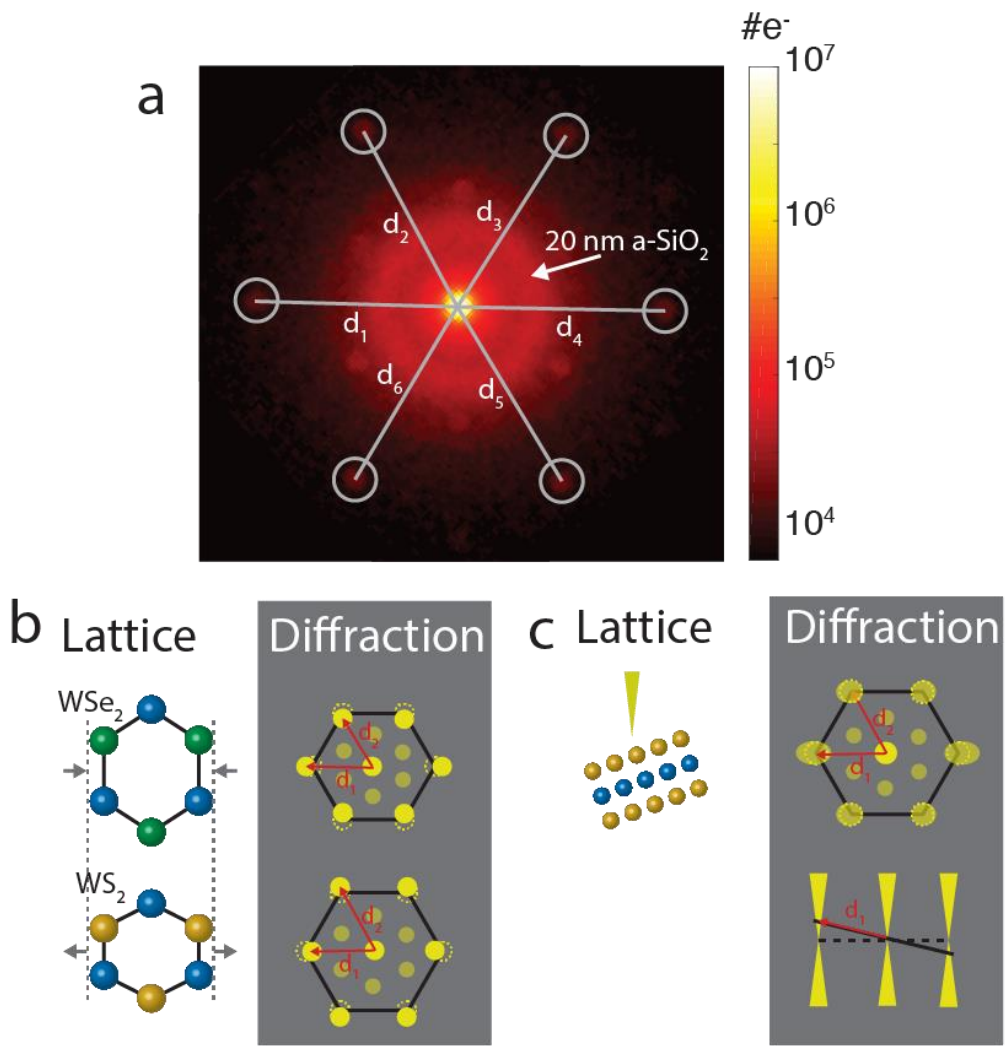

**Figure S3 | Lattice constant map. a,** Diffraction pattern of $WS_2$-$WSe_2$ on 20 nm $SiN_x$ windows. The gray circles are the masks we used to calculate the CoMs. The reciprocal lattice constants were measured, shown as $d_1$ – $d_6$. **b,** Schematic showing how strain affects the lattice constant measurements. **c,** Schematic showing how small tilt affects the lattice constant measurements. The small strain and tilt are higher order effect for lattice constant calculation.

## 3: Mapping strain

To map the strain from the 4D dataset, we calculated the diffraction vectors $g_{i\ (i=1,2)}$ (i.e. the reciprocal lattice vectors) as shown in Fig. S4a. The reference diffraction vectors $g_i^{ref}$ were set by averaging 200 scan positions (or pixels) in real space where half of them are on WS$_2$ and the other half are on the WSe$_2$ region. The choice of these $g_i^{ref}$ is for mapping convenience. Afterwards, we derived the transformation matrix T using $g_i$ = T $g_i^{ref}$. T can be polar-decomposed into a rotation matrix R and a strain matrix U, from which the uniaxial strain $\varepsilon_{xx}$=1-U$_{11}$ and $\varepsilon_{yy}$=1-U$_{22}$, shear strain $\varepsilon_{xy}$=U$_{12}$, and rotation $\varepsilon_{rot}$=asin(R$_{12}$) can be calculated.

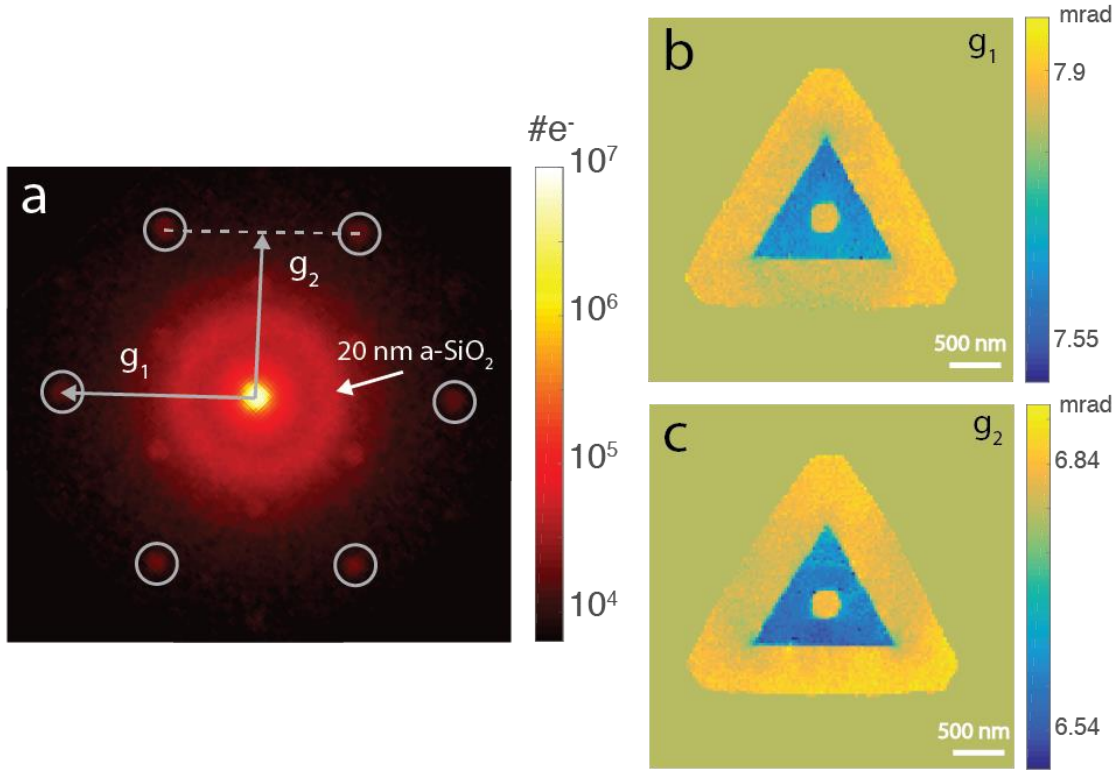

**Figure S4 | Maps of diffraction vectors. a,** A single diffraction pattern of WSe$_2$ taken by EMPAD with second order diffraction spots highlighted by the masks. For each spot, we calculated the CoM and achieve the diffraction vectors as labeled by $g_1$ and $g_2$. **b,c,** |$g_1$| and |$g_2$| maps over the entire triangle.

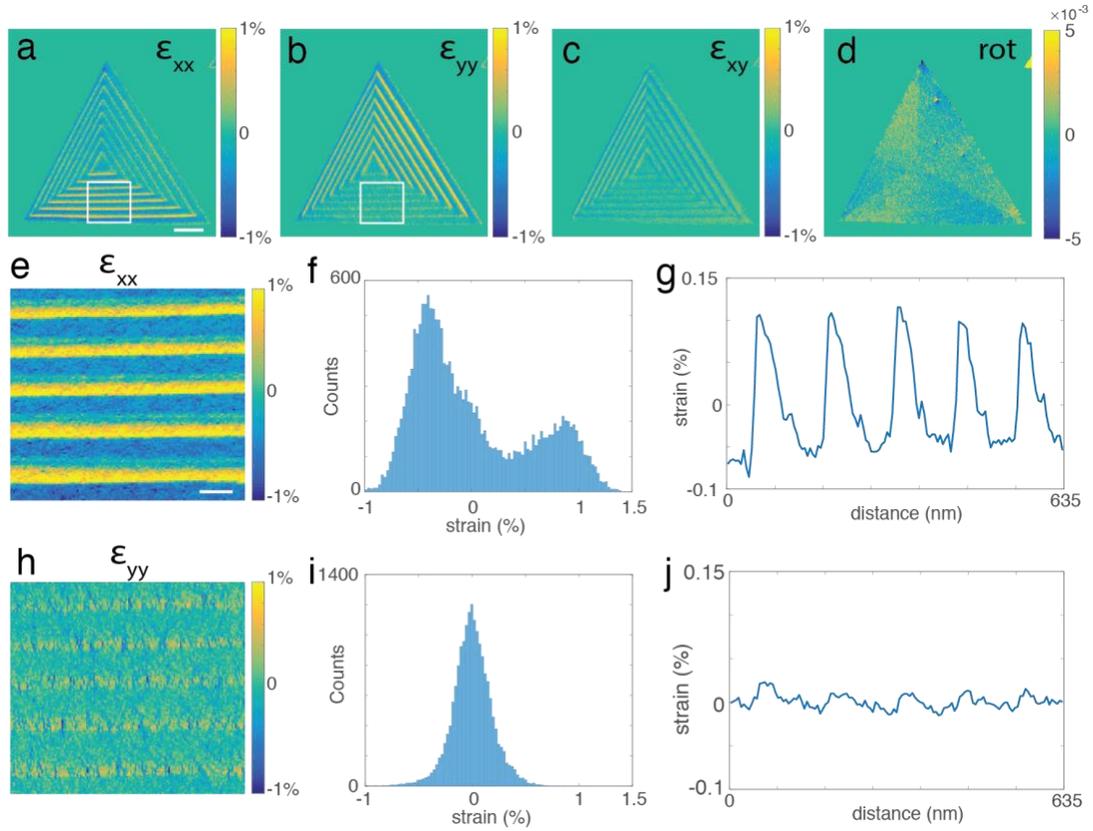

**Figure S5 | Strain maps of WS$_2$-WSe$_2$ superlattices. a-d,** x-direction uniaxial strain (a), y-direction uniaxial strain (b), shear strain (c), and rotation (d) maps. **e,h,** Magnified x-direction uniaxial strain and y-direction uniaxial strain maps from the white boxes in (a) and (b). **f,g,** Histogram and the line profiles of (e). **i,j,** Histogram and line profiles of (h).

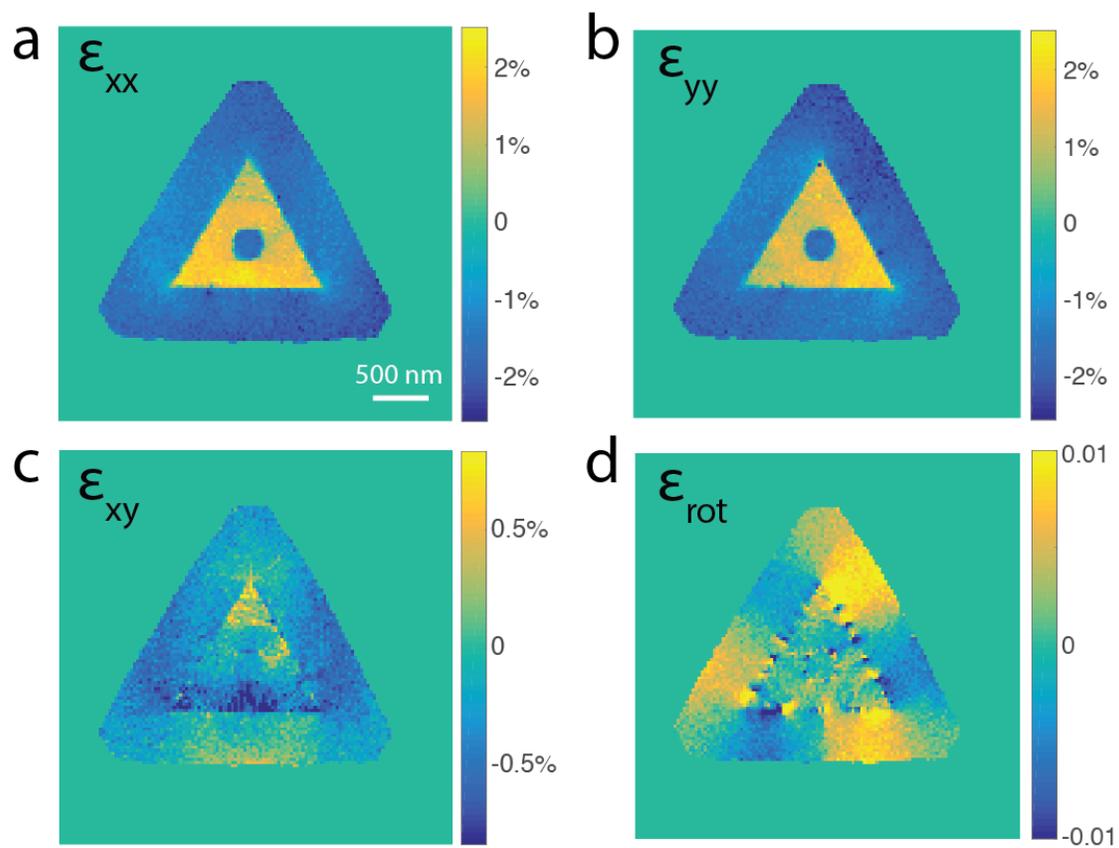

**Figure S6 | Strain maps of thick WS$_2$-WSe$_2$ junctions. a**, x-direction uniaxial strain map. **b,** y-direction uniaxial strain map. **c,** shear strain map. **d,** rotation map.

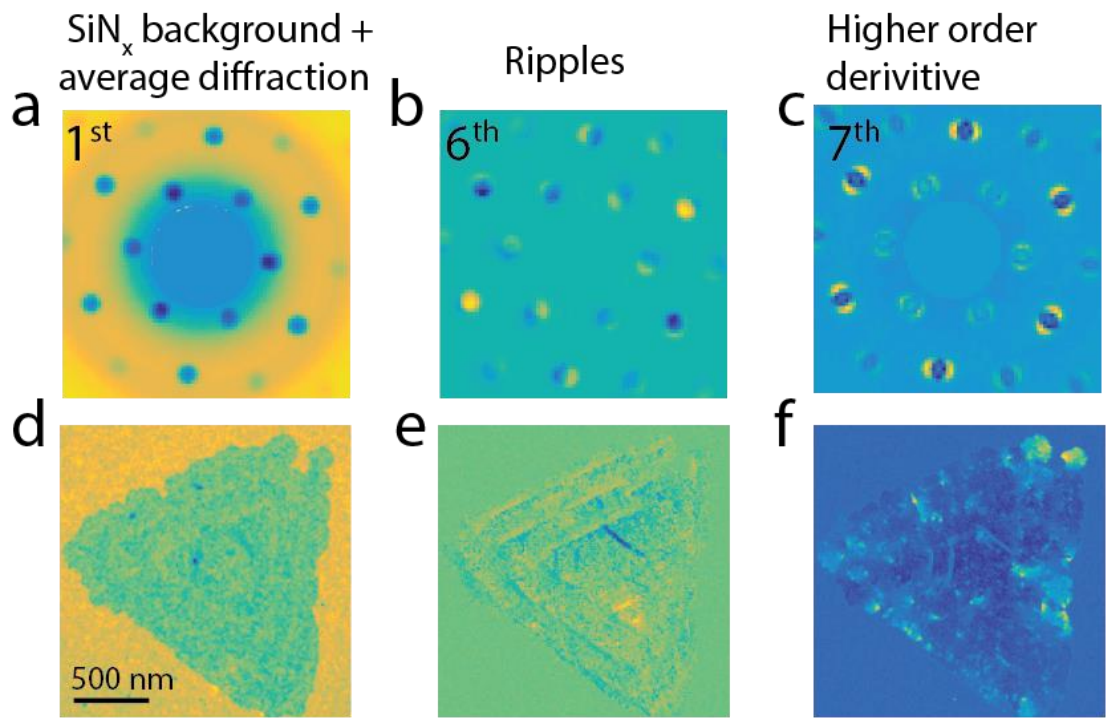

**Figure S7 | PCA of 4D dataset. a,d,** The first principle component which represents the SiN$_x$ substrate and the averaged diffraction patterns from the lattice. **b,e,** The sixth principle component representing the ripple with a different orientation from the one in Fig. 5i and j. **c,f,** The seventh principle component showing higher order derivatives of the diffraction patterns. For principle components more than seventh order, the real space images start to show less feature as the higher order terms mainly contain noise.